\documentclass[aps,prb,twocolumn,showpacs]{revtex4-1} 
\usepackage{array}
\usepackage{amsbsy}
\usepackage{bm}
\usepackage{color}
\usepackage{graphicx}
\usepackage[hyperindex,breaklinks,linktocpage,colorlinks]{hyperref}   

\def\d#1{{\rm d}#1}
\def\eps{\varepsilon}

\def\gdir#1{\langle #1 \rangle}

\def\refeq#1{(\ref{#1})}
\def\reffig#1{Fig.~\ref{#1}}
\def\reffigs#1{Figs.~\ref{#1}}

\def\sym{{\rm sym\ }}

\begin{document}
\title{On the incompatibility of strains and its application to mesoscopic studies of plasticity}

\author{R. Gr\"oger}
\email{groger@ipm.cz}
\affiliation{Institute of Physics of Materials, Academy of Sciences of the Czech Republic, 
\v{Z}i\v{z}kova 22, Brno 616 62, Czech Republic}

\author{T. Lookman}
\author{A. Saxena}
\affiliation{Theoretical Division, Los Alamos National Laboratory, Los Alamos, NM 87545, USA}

\begin{abstract}
  Structural transitions are invariably affected by lattice distortions. If the body is to remain
  crack-free, the strain field cannot be arbitrary but has to satisfy the Saint-Venant compatibility
  constraint. Equivalently, an incompatibility constraint consistent with the actual dislocation
  network has to be satisfied in media with dislocations. This constraint can be incorporated into
  strain-based free energy functionals to study the influence of dislocations on phase stability. We
  provide a systematic analysis of this constraint in 3D and show how three incompatibility equations
  accommodate an arbitrary dislocation density. This approach allows the internal stress field to be
  calculated for an anisotropic material with spatially inhomogeneous microstructure and
  distribution of dislocations by minimizing the free energy. This is illustrated by calculating the
  stress field of an edge dislocation and comparing it with that of an edge dislocation in an
  infinite isotropic medium. We outline how this procedure can be utilized to study the interaction
  of plasticity with polarization and magnetization.
\end{abstract}

\date{\today}



\maketitle


\section{Introduction}

Physical properties of solids are mainly determined by crystal defects. Their strain fields couple
to the microstructure and thus often deteriorate the properties of the underlying perfect
crystal. The coupling of point defects with microstructure can give rise to variation in the twin
wall widths of ferroelastic martensites \cite{lee:05}. Similarly, dislocations through their
long-range strain fields affect the spatial evolution of the microstructure and are typically
responsible for shifts of the transformation temperature \cite{sato:82, miyazaki:86, groger:08}. To
guarantee phase stability, one needs to understand how the phase diagram changes as a function of
the density of dislocations. This can be accomplished efficiently using the continuum theory of
dislocations \cite{kroner:58, kroupa:62, kosevich:79} in which dislocations are viewed as sources of
incompatibility of strains and stresses \cite{teodosiu:82, kleinert:89}. 

The concept of incompatibility of strains was used as a basis for the development of field
dislocation mechanics \cite{acharya:01} and continues to serve as a template for the development of
mesoscopic models of self-organization of dislocation networks \cite{bako:99, el-azab:00, zaiser:01,
  limkumnerd:06}. It has been readily incorporated within the Landau continuum approach to study
spinodal decomposition mediated by dislocations \cite{haataja:04} as well as the effects of dislocations
on the stability of phases in martensites \cite{groger:08}. These studies have been in two
dimensions where only one incompatibility relation constrains the spatial variation of order
parameter fields \cite{kerr:99, lookman:03}. Analyses of the three-dimensional compatibility
constraint, which guarantees integrability of the strain field in dislocation-free media, can be
found in \cite{malyi:86, borg:90, selvadurai:00}.  In three dimensions, the compatibility constraint
is represented by six equations whereas only three are required to make the strain field integrable
\cite{rasmussen:01}. It has not been apparent how this reduction can be justified and this lack of
  understanding has led to conflicting statements about the mutual dependencies of these equations
  \cite{andrianov:03}. Nevertheless, a systematic analysis that examines the structure of the
three-dimensional incompatibility constraint as applied to a finite density of dislocations, is
lacking. This has not allowed for the generalization of two-dimensional models of plasticity to
three dimensions. The primary objective of this paper is to provide a systematic analysis of the
full set of six incompatibility equations assuming periodic boundary conditions and to show how they
can be reduced to three incompatibility constraints that represent an arbitrary dislocation
network. This makes it possible to develop strain-based free energies \cite{barsch:84, kartha:95} in 3D
that describe the spatial evolution of the microstructure at the length and time scales that are not
amenable to atomistic or continuum approaches.

The elastic free energy due to a distortion from a high symmetry parent phase to a lower symmetry
product phase is typically written in terms of a finite number of order parameters that characterize
this change.  Thus, the transformation from a cubic symmetry to the tetragonal, orthorhombic or
monoclinic symmetries involves symmetry-allowed invariants in the free energy, the \emph{mathematical
  structure} of which is the same for all materials that undergo the same type of phase
transition. Materials undergoing the same crystal symmetry change are distinguished by the
coefficients of these invariants that reflect the temperature and pressure dependencies of elastic
constants, martensitic strain at the transition or curvature of soft phonon modes that drive the
transition. This is extremely valuable since a single formulation of the free energy then describes
a large class of materials. In this paper, we consider the strain-based formulation of the
Landau-Ginzburg functional \cite{landau:86} for the cubic to tetragonal transformation that was
developed by Barsch and Krumhansl \cite{barsch:84} and successfully applied in a number of
theoretical studies \cite{bishop:03, ahluwalia:06, groger:08}.

\section{Incompatibility constraint}

We begin by defining a set of six symmetry-allowed combinations of the components of the elastic
strain tensor which can serve as order parameters to characterize the symmetry change from the
parent (typically cubic) phase to an equal or lower symmetry phase. The free energy is then written
in terms of these order parameters and its minimization yields the homogeneous ground state as well
as spatial variations in the order parameter fields due to any inhomogeneities, such as gradient
terms. However, in order to keep the deformed body continuous (i.e. without cracks), the components
of the total distortion tensor $\bm{\beta}^t = \bm{\eps}^t + \bm{\omega}^t$ cannot be independent
($\bm{\eps}^t$ is the symmetric elastic-plastic strain tensor and $\bm{\omega}^t$ the antisymmetric
tensor of elastic-plastic rotations). Instead, they satisfy the equation \cite{kroner:58} $\nabla
\times \bm{\beta}^t = \bm{0}$, where the total distortion is $\bm{\beta}^t = \bm{\beta} +
\bm{\beta}^p$. Here, $\bm{\beta}^p$ is the permanent plastic distortion of the lattice due to the
spatially inhomogeneous distribution of dislocations. It is then customary to write $\nabla \times
\bm{\beta}^p = -\bm{\alpha}$, where $\bm{\alpha}$ is the Nye tensor that represents the density of
infinitesimal Burgers vectors \cite{nye:53}. However, the cohesive forces of the matter cause
elastic relaxation of the lattice, $\bm{\beta}$. In order for the total distortion tensor to be
curl-free, it follows that $\nabla \times \bm{\beta} = \bm{\alpha}$. This tensor is non-symmetric
and its components, $\alpha_{ij}$, correspond to a dislocation whose line direction is parallel to
$x_i$ and the Burgers vector parallel to $x_j$. Hence, the diagonal components of $\bm{\alpha}$
correspond to the three screw dislocations with their line directions parallel to the three $\langle
100 \rangle$ axes, while the six off-diagonal components describe the density of the six variants of
edge dislocations.  Taking the symmetric part of the curl of $\nabla \times \bm{\beta} =
\bm{\alpha}$ yields the so-called incompatibility constraint
\begin{equation}
  \nabla \times \nabla \times \bm{\eps} = \bm{\eta} \ ,
  \label{eq:comp}
\end{equation}
where $\bm{\eta} = \sym (\nabla \times \bm{\alpha})$ is the incompatibility tensor.  

When developing models of phase transformations based on strains, the condition \refeq{eq:comp} has
to be enforced \cite{kroner:58}. In components, Eq.~\refeq{eq:comp} represents two sets of three
equations,
\begin{eqnarray} 
  \nonumber
  \eps_{kk,jj} - 2\eps_{jk,jk} + \eps_{jj,kk} &=& \eta_{ii} \\ 
  \eps_{ik,jk} - \eps_{kk,ij} - \eps_{ij,kk} + \eps_{jk,ik} &=& \eta_{ij} \ ,
  \label{eq:incomp2}
\end{eqnarray}
where $i,j,k=\{1,2,3\}$ and $i\not=j\not=k$. The comma means a partial derivative and no summation
of the repeated indices is performed here. In two dimensions, only one equation is not satisfied
trivially and this represents the constraint for the internal strains. The situation is more
complicated in three dimensions since all six equations have to be enforced. However, only three
equations are needed and the main problem is to find a rigorous way to reduce the six equations to three.

In order to find the minimum number of constraints for a spatial variation of the internal elastic
strain field that complies with a given dislocation density, assuming periodic boundary
conditions, we begin by writing  \refeq{eq:incomp2} in k-space. In the augmented matrix
representation, this linear system takes on a block-symmetric form
\begin{equation}
  \small
  \left(
  \begin{array}{cccccc}
    0 & -k_3^2 & -k_2^2 & 2k_2k_3 & 0 & 0 \\
    -k_3^2 & 0 & -k_1^2 & 0 & 2k_1k_3 & 0 \\
    -k_2^2 & -k_1^2 & 0 & 0 & 0 & 2k_1k_2 \\
    k_2k_3 & 0 & 0 & k_1^2 & -k_1k_2 & -k_1k_3 \\
    0 & k_1k_3 & 0 & -k_1k_2 & k_2^2 & -k_2k_3 \\
    0 & 0 & k_1k_2 & -k_1k_3 & -k_2k_3 & k_3^2
  \end{array}
  \left|
  \begin{array}{c}
    \tilde\eta_{11} \\ \tilde\eta_{22} \\ \tilde\eta_{33} \\
    \tilde\eta_{23} \\ \tilde\eta_{13} \\ \tilde\eta_{12}
  \end{array}
  \right.
  \right) \ ,
  \label{eq:incomp_mat}
\end{equation}
where tilde ($\tilde{\ }$) denotes a Fourier image, and $\bm{k}=(k_1,k_2,k_3)$ is the k-space
vector. If no additional constraints on the components $\tilde\eta_{ij}$ are supplied, this system
of equations would be internally inconsistent. The three conditions that make this system consistent
are obtained from the Bianchi identity $\nabla \cdot \bm{\alpha} = \bm{0}$ which is only true for a
permissible dislocation density, i.e. if dislocations do not begin or end inside the body. In
order to gain insight into the structure of \refeq{eq:incomp_mat}, let us begin by splitting the
system into two sets of three equations each, where the first set is obtained from
\refeq{eq:incomp_mat} by taking rows number 1-3 and the second set from rows number 4-6. Performing
the row reduction on each of these subsystems separately converts them into the following forms:
\begin{eqnarray}
  \nonumber
  \bm{M} \tilde{\bm{\eps}}^{\rm edge} &=& \tilde{\bm{\eta}}^{\rm edge} \\
  \bm{M} \tilde{\bm{\eps}}^{\rm screw} &=& \tilde{\bm{\eta}}^{\rm screw} \ ,
  \label{eq:2subsys}
\end{eqnarray}
where the matrices of both systems are given by
\begin{equation}
  \bm{M} = 
  \left(
  \begin{array}{cccccc}
    1 & 0 & 0 & \frac{k_1^2}{k_2k_3} & -\frac{k_1}{k_3} & -\frac{k_1}{k_2} \\
    0 & 1 & 0 & -\frac{k_2}{k_3} & \frac{k_2^2}{k_1k_3} & -\frac{k_2}{k_1} \\
    0 & 0 & 1 & -\frac{k_3}{k_2} & -\frac{k_3}{k_1} & \frac{k_3^2}{k_1k_2}
  \end{array}
  \right) \ .
\end{equation}
However, the right-hand sides of \refeq{eq:2subsys} are different:
\begin{equation}
  \tilde{\bm{\eta}}^{\rm edge} = 
  \left(
  \begin{array}{cccccc}
    \frac{k_1^2\tilde\eta_{11}-k_2^2\tilde\eta_{22}-k_3^2\tilde\eta_{33}}{2k_2^2k_3^2} \\
    \frac{-k_1^2\tilde\eta_{11}+k_2^2\tilde\eta_{22}-k_3^2\tilde\eta_{33}}{2k_1^2k_3^2} \\
    \frac{-k_1^2\tilde\eta_{11}-k_2^2\tilde\eta_{22}+k_3^2\tilde\eta_{33}}{2k_1^2k_2^2}
  \end{array} \right) , \quad
  \tilde{\bm{\eta}}^{\rm screw} = 
  \left(
  \begin{array}{cccccc}
    \frac{\tilde\eta_{23}}{k_2k_3} \\
    \frac{\tilde\eta_{13}}{k_1k_3} \\
    \frac{\tilde\eta_{12}}{k_1k_2}
  \end{array} \right) .
\end{equation}
It is important to emphasize that the right-hand side of the first system contains only the diagonal
components of $\tilde{\bm{\eta}}$. If an edge dislocation is present, only
$\tilde{\bm{\eta}}^{\rm edge}$ is nonzero whereas  $\tilde{\bm{\eta}}^{\rm screw} = \bm{0}$. Similarly, the
right-hand side of the second system contains only the off-diagonal components of
$\tilde{\bm{\eta}}$. Hence, for a screw dislocation $\tilde{\bm{\eta}}^{\rm screw}$ is nonzero whereas
$\tilde{\bm{\eta}}^{\rm edge} = \bm{0}$. Clearly, the first set of equations can be used to incorporate
edge dislocations, while the second set incorporates screw dislocations.  However, one cannot solve
these systems to obtain the strain field in terms of the components of the incompatibility tensor
$\tilde{\bm{\eta}}$. They only represent constraints that the internal strain field has to satisfy
if a finite density of dislocations is present. It is also very important to emphasize that since
the right-hand sides of these systems are different, the constraints imposed on strains by the two
systems are in general different. Hence, we denote by $\tilde{\bm{\eps}}^{\rm edge}$ and
$\tilde{\bm{\eps}}^{\rm screw}$ the total elastic strain fields due to \emph{all} edge and screw dislocations,
respectively. Because the two systems are linear, the superposition of the two sets yields
\begin{equation}
  \bm{M}\tilde{\bm{\eps}} = \tilde{\bm{\eta}}^{\rm edge} + \tilde{\bm{\eta}}^{\rm screw}\ ,
  \label{eq:incomp3}
\end{equation}
where $\tilde{\bm{\eps}} = \tilde{\bm{\eps}}^{\rm edge} + \tilde{\bm{\eps}}^{\rm screw}$ is the
Fourier image of the total elastic strain tensor. Eq. \refeq{eq:incomp3} represents a linear system
of three incompatibility constraints whose right-hand side contains in general all components of the
incompatibility tensor. Hence, in three dimensions \refeq{eq:incomp3} are three conditions the
internal strain field has to satisfy due to the existence of an arbitrary distribution of the
dislocation density. If these equations are written as constraints, i.e. $\tilde{g}_i[\tilde{\bm{\eps}}] =
0$ and $i=\{1,2,3\}$, one can write the free energy density for a ferroelastic material in k-space
as
\begin{equation}
  \tilde{F}[\tilde{\bm{\eps}}] = \int_\Omega \left\{ \tilde{f}[\tilde{\bm{\eps}}] + \sum_i \lambda_i
  \tilde{g}_i[\tilde{\bm{\eps}}] \right\} \d{\bm{k}} \ ,
  \label{eq:freeE}
\end{equation}
where $\tilde{f}[\tilde{\bm{\eps}}]$ is the k-space expression of the free energy density,
$\lambda_i$ the Lagrange multipliers, and $\Omega$ the volume of the k-space. Recall that for
two-dimensional problems, the sum in \refeq{eq:freeE} contains only one term, namely that
corresponding to $\tilde{\eta}_{ii}$, where $x_i$ is the direction of the (straight)
dislocation line. However, in three dimensions, this sum contains three terms that impose the three
conditions embodied in \refeq{eq:incomp3}. The development above is completely general, applicable
to any kind of displacive phase transformation and an arbitrary (but permissible) dislocation
density. Below, we use this formulation to outline the development of a self-consistent mesoscopic
model for a cubic to tetragonal transformation in which the microstructure is coupled to the density
of dislocations.

\section{Cubic to tetragonal transition}

We begin by defining a set of six order parameters $e_i[\bm{\eps}]$ ($i=1,\ldots,6$), one of which
corresponds to a hydrostatic distortion, two to the change of shape from cubic to tetragonal, and
three are shear components. The concrete expressions for these order parameters can be found in
\cite{barsch:84}. Here, the primary order parameters are those that represent the tetragonal
distortion of the lattice, while all others are secondary order parameters. If one considers the
cubic structure as a reference phase of zero energy, then the strain energy density of a distorted
phase relative to this reference structure can be written as $f[\bm{\eps}] =
\frac{1}{2}c_{ijkl}\eps_{ij}\eps_{kl}$.  Since this is harmonic, its minimum corresponds to the
reference cubic structure. In order to allow for a phase transition to occur, one augments the free
energy by higher order terms of the primary order parameters that respect the cubic
symmetry. Finally, the length scale is included by adding gradient terms of the primary order
parameters. The free energy density is then written as $f[\bm{e}]$, where $\bm{e}$ is a vector of
the six order parameters. Similarly, we can obtain an equivalent expression for the three
incompatibility conditions in terms of the order parameters. The stationary order parameter fields
corresponding to a fixed dislocation density (i.e. incompatibility field $\bm{\eta}$), are then
obtained by variational minimization of the free energy equivalent to \refeq{eq:freeE}. Since the
components of the elastic strain tensor are defined in terms of the order parameters, the internal
elastic strain field is straightforward to calculate. The conjugate stress field is then obtained at
once using the Hooke's law.

\begin{figure*}
  \centering
  \includegraphics[scale=0.16]{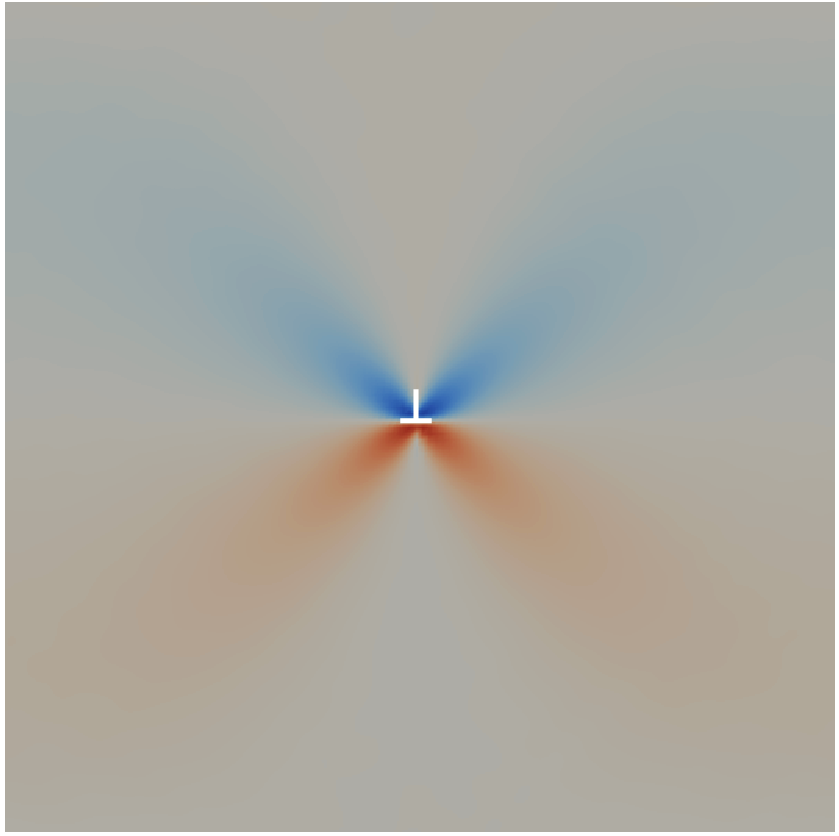} \quad
  \includegraphics[scale=0.25]{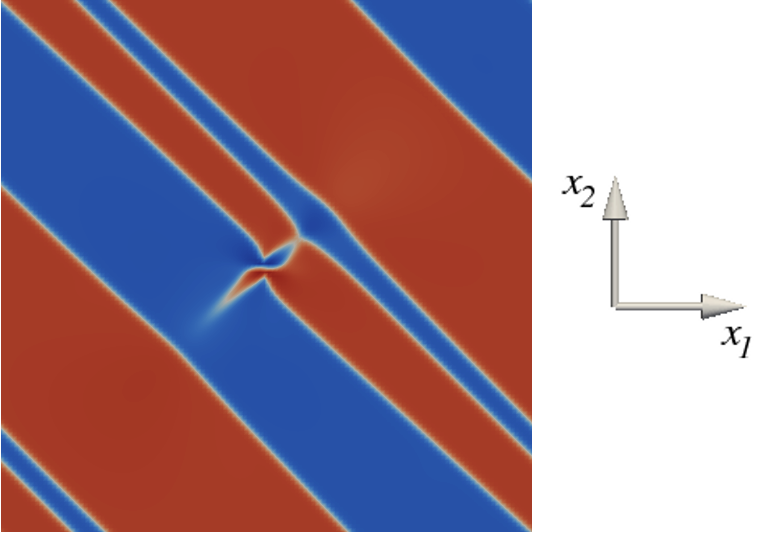} \\
  \hskip-2cm (a) \hskip4.5cm (b) \\[1em]
  \includegraphics[scale=0.25]{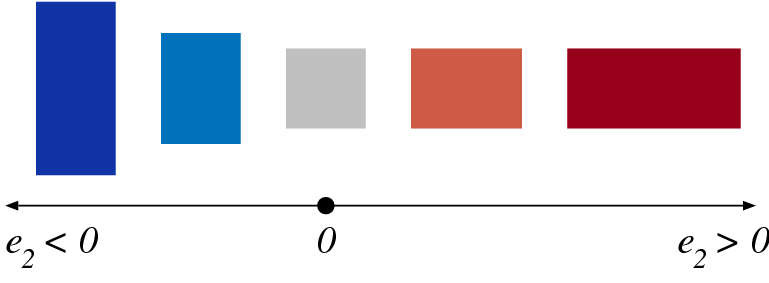} \\
  (c)
  \caption{Characteristic deviatoric distortion ($e_2$) of the lattice caused by an edge dislocation
    in the middle of the block above (a) and below (b) the transformation temperature in a periodic
    anisotropic medium.  As shown in (c), blue ($e_2<0$) and red ($e_2>0$) are
    the two characteristic tetragonal distortions and gray corresponds to the parent cubic
    lattice.}
  \label{fig:e2}
\end{figure*}

Our objective in the following is to demonstrate how the internal strain field is disturbed when a
single edge dislocation is present. It can be shown that without dislocations, the minimum of the
free energy corresponds to a twinned microstructure with the normal parallel to the $\gdir{110}$
direction. Due to the rotational symmetry of the parent cubic lattice, we may consider that the
normal of the martensite plate is parallel to $[110]$. If we further consider a straight edge
dislocation whose line direction is parallel to $x_3$, all fields are functions of $x_1$ and $x_2$
but not of $x_3$. This allows us to reduce the problem to the two dimensional plane strain case.

Let us now insert a $[100]$ edge dislocation into the middle of the block, as shown in
\reffig{fig:e2}a. Here, the Burgers vector of the dislocation is taken artificially as the edge
length of one unit cell used to discretize the space in our simulation.  Due to the periodic
boundary conditions, this represents an array of straight edge dislocations whose line directions
are parallel to $x_3$ and the Burgers vectors point in the positive $x_1$ direction. Hence, the only
nonzero component of the Nye tensor is $\alpha_{31}$ and the only nonzero component of the
incompatibility tensor is $\eta_{33}$. This incompatibility is incorporated into the free energy by
the procedure explained above. Minimization of the free energy then yields the equilibrium internal
strain field. This is represented using the order parameters $e_i$. Since we are interested in the
phase transformation from the parent cubic phase to the product tetragonal phase, we will be
concerned with the spatial variation of $e_2$, which is proportional to the deviatoric strain
$\eps_{11}-\eps_{22}$. This minimization was first performed above $T_c$ and \reffig{fig:e2}a shows
the calculated spatial variation of the order parameter field $e_2$. If the dislocation was not
present, the minimum of the free energy would correspond to the cubic phase (gray color). However,
if the dislocation is present, the spatially homogeneous microstructure is frustrated by the strain
field of the dislocation. One can imagine that the dislocation was created by sliding into the block
an extra half-plane $(x_2,x_3)$ at $x_2>0$ which terminates in the middle of the block. Hence, the
surrounding lattice at $x_2>0$ is under compression in the $x_1$ direction. Because $\eps_{11}<0$
and $\eps_{22} \approx 0$, the order parameter $e_2$ becomes negative and this corresponds to the
blue region in \reffig{fig:e2}a. Due to the termination of the extra half-plane at $x_2=0$, the
lattice at $x_2<0$ is under tension in the $x_1$ direction. Since $\eps_{11}>0$ and $\eps_{22}
\approx 0$ the order parameter $e_2$ is positive here and this gives rise to the red part of the
butterfly shape in \reffig{fig:e2}a. If the minimization of the free energy was performed below
$T_c$ and the body was free of dislocations, the equilibrium field $e_2$ would represent a twinned
microstructure in which the two tetragonal variants would be separated by twin boundaries of the
cubic symmetry. As shown in \reffig{fig:e2}b, the presence of the dislocation causes a local
distortion of this twinned microstructure. This results in a misorientation of the direction of the
tetragonal distortion and an emergence of curvature, both in the neighborhood of the dislocation and
also farther away where the tip of the distortion meets another twin boundary. In addition to this
distortion, the microstructure accommodates the dislocation as a part of the twin boundary. The
distortion around the dislocation thus represents signatures of the dislocation in the order
parameter field $e_2$ above and below $T_c$. For distances large compared to the unit cell in the
calculation, both fields in \reffigs{fig:e2}a,b approach those for the dislocation-free body above
and below $T_c$, respectively.

\begin{figure*}
  \centering
  \includegraphics[scale=0.25]{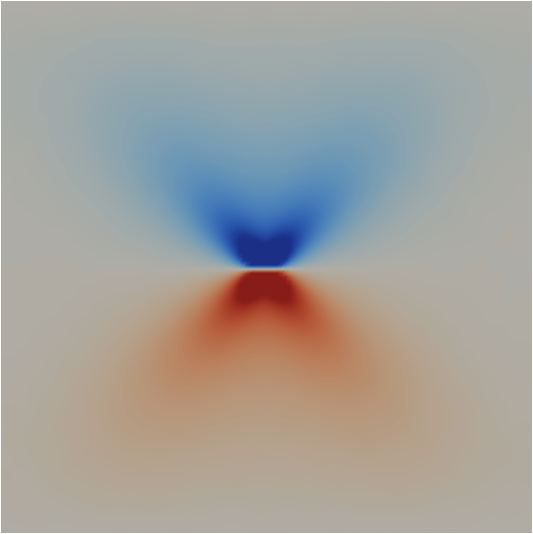} \quad
  \includegraphics[scale=0.25]{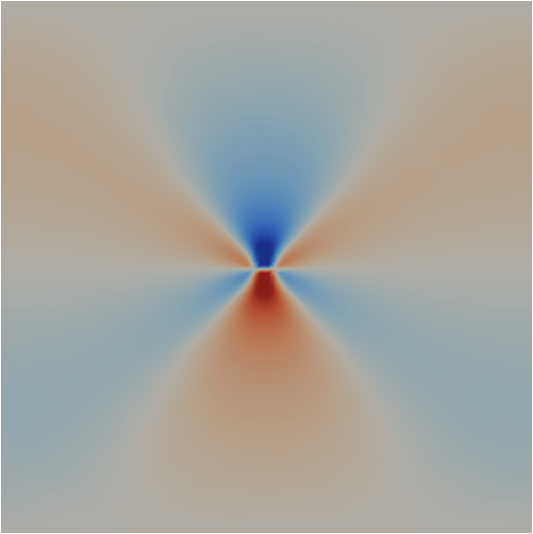} \quad
  \includegraphics[scale=0.25]{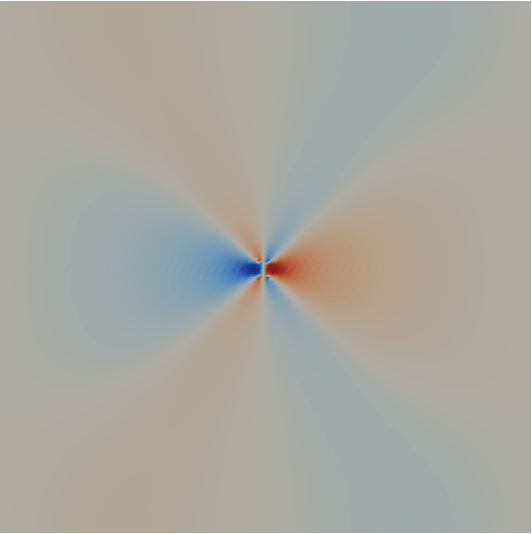} \\
  (a) \hskip4.5cm (b) \hskip4.5cm (c) \\[1em]
  \includegraphics[scale=0.25]{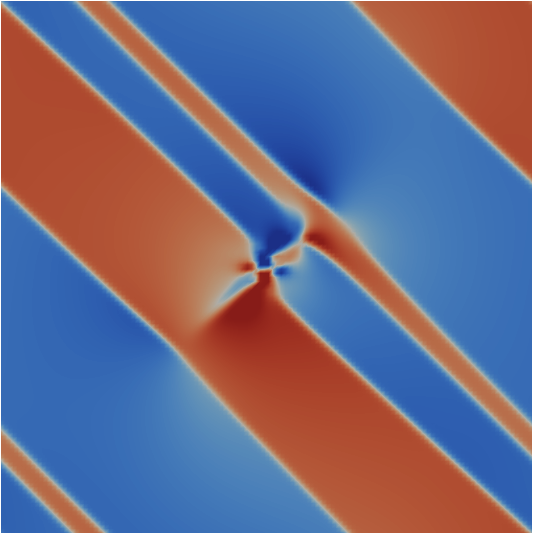} \quad
  \includegraphics[scale=0.25]{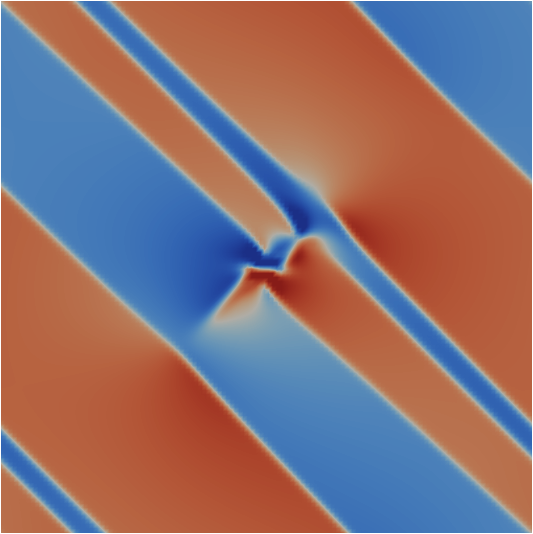} \quad
  \includegraphics[scale=0.25]{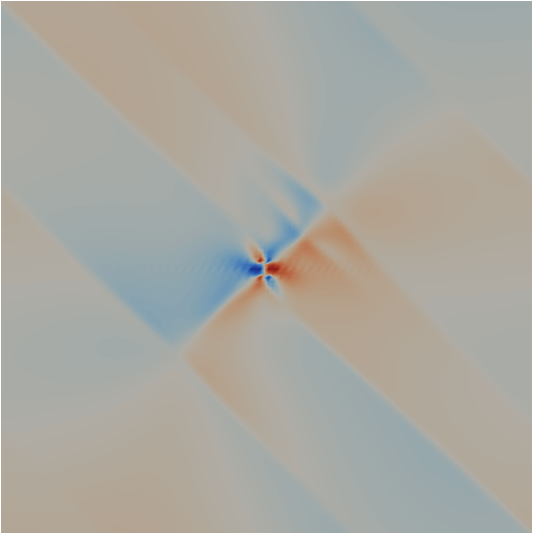} \\
  (d) \hskip4.5cm (e) \hskip4.5cm (f) \\
  \caption{Calculated stress field around an edge dislocation in a periodic anisotropic medium 
    (a-c) above and (d-f) below $T_c$. Here, (a) and (d) corresponds to $\sigma_{11}$, (b) and (e) to
    $\sigma_{22}$, and (c) and (f) to $\sigma_{12}$. The color scale ranges from negative values
    (dark blue), via zero (gray), to positive values (dark red). In all figures the color map spans
    the same range of magnitudes of stresses.}
  \label{fig:sig}
\end{figure*}

For completeness, we show in \reffigs{fig:sig}a-c and \reffigs{fig:sig}d-f the three components of
the calculated internal stress field above and below $T_c$, respectively. Here, red represents
positive stresses, gray the unstressed regions, and blue negative stresses. Above $T_c$, the stress
field in \reffigs{fig:sig}a-c agrees qualitatively with that obtained from analytical expressions of
the stress field around an edge dislocation in an infinite isotropic medium\cite{hirth:82}. Below
$T_c$, the stress field in \reffigs{fig:sig}d-f includes nonlocal interactions of the the stress field
of the dislocation with the microstructure. No analytical expression for the stress field exists due
to the complexity of the problem. As we showed already in \reffig{fig:e2}b, an edge dislocation
below $T_c$ is responsible for developing curvature in the system. This shows up in the stress field
by local bending of the twin walls in the vicinity of the dislocation and reversals of the state of
stress in the direction perpendicular to the twin wall.

Although the calculation above has been performed for one edge dislocation, this approach is
completely general in that the right-hand side of \refeq{eq:incomp3} can represent an arbitrary
dislocation network.

\section{Generalization}

The formalism developed above can be systematically extended to study the influence of dislocations
on the physical properties of multiferroics. To describe piezoelectric/piezomagnetic materials in
which polarization/magnetization is coupled to strain \cite{ahluwalia:05}, one typically begins by
writing the Gibbs free energy density
\begin{equation}
  f[\bm{q},\bm{\eps}] = f_{\rm Landau}[\bm{q}] + \frac{1}{2}\sum_{ijkl}c_{ijkl}\eps^s_{ij}\eps^s_{kl} + \sum_i \lambda_i
  g_i[\bm{\eps}] \ ,
  \label{eq:fpiezo}
\end{equation}
where $\bm{q}$ is a vector of order parameters and $f_{\rm Landau}[\bm{q}]$ the Landau free energy
density consistent with the symmetry of the underlying crystal structure. The second term in
\refeq{eq:fpiezo} represents the elastic strain energy density and the third term incorporates the
(in)compatibility constraint derived earlier. The spontaneous strain $\bm{\eps}^s$ is written as
$\bm{\eps}^s = \bm{\eps} - \bm{\eps}^{\bm{q}}$, where $\bm{\eps}$ is the total elastic strain and
$\eps_{ij}^{\bm{q}} = Q_{ijkl}q_kq_l$ the strain caused by the change of the order parameter. For
piezoelectric materials, $\bm{Q}$ represents the electrostrictive tensor and $\bm{q}$ the
polarization vector $\bm{P}$, while in piezomagnetic materials $\bm{Q}$ is the magnetostrictive
tensor and $\bm{q}$ the magnetization vector $\bm{M}$. The first term in \refeq{eq:fpiezo} then
depends only on polarization (magnetization), the second term includes a bilinear coupling of the
total elastic strain with polarization (magnetization), and the last term represents a constraint
for the spatial variation of strain through which dislocations affect the order parameter field
$\bm{q}$. The stationary state of the system is thus represented by a generally nontrivial field of
polarization or magnetization, $Q_i[Q_{j\,(j\not=i)},\bm{\alpha}]$, where $\bm{\alpha}$ is the Nye
tensor of the dislocation density.

\section{Conclusion}

The main result of this paper is a set of three incompatibility constraints \refeq{eq:incomp3} that
provide a constraint for the internal strain field in an elastic medium with dislocations. For the
special case of zero dislocation density, i.e. by setting the right-hand side of \refeq{eq:incomp3}
to zero, one directly obtains the three compatibility equations that have to be satisfied at any
point in media without dislocations. We have shown how the constraints \refeq{eq:incomp3} can be
systematically incorporated into the strain-based free energy functional to obtain a mesoscopic
description of phase transitions mediated by dislocations. Since the lattice constitutes an elastic
template that is common to all materials, this formalism can be applied to couple dislocations not
only with microstructure but also with other properties of functional materials such as
polarization and magnetization.

\begin{acknowledgments}
  This research was supported by the Marie-Curie International Reintegration Grant No. 247705
  ``MesoPhysDef'', and in part by the Academy of Sciences of the Czech Republic (Research Project
  No. AV0Z20410507) and by the U.S. Department of Energy.
\end{acknowledgments}

\bibliography{bibliography}

\end{document}